\documentclass[10pt,twocolumn]{article}

\usepackage[margin=0.75in]{geometry}
\usepackage{amsmath}
\usepackage{booktabs}
\usepackage{graphicx}
\usepackage{hyperref}
\usepackage{authblk}
\usepackage{enumitem}
\usepackage[numbers]{natbib}
\usepackage{caption}
\usepackage{float}

\begin{document}

\title{Preserving Disagreement: Architectural Heterogeneity and\\Coherence Validation in Multi-Agent Policy Simulation}

\author{Ariel Sela}
\affil{Tel Aviv University \\ \texttt{arielsela1@mail.tau.ac.il}}

\maketitle

\begin{abstract}
Multi-agent deliberation systems using large language models (LLMs) are increasingly proposed for policy simulation, yet they suffer from a failure mode specific to normative deliberation: artificial consensus, where evaluator agents converge on the same option regardless of their assigned value perspectives. Unlike accuracy-oriented tasks where convergence signals correctness, in policy deliberation convergence obscures the genuine trade-offs that make the output useful. We present the AI Council, a three-phase deliberation framework (structured debate, independent evaluation, coherence validation) and conduct 120 deliberations across two policy scenarios to test two interventions against this problem. First, architectural heterogeneity (assigning a different 7--9B parameter model to each value perspective) significantly reduces first-choice concentration compared to a homogeneous single-model baseline (child welfare: 70.9\% $\rightarrow$ 46.1\%, $p < 0.001$, $r = 0.58$; housing: 46.0\% $\rightarrow$ 22.9\%, $p < 0.001$, $r = 0.50$). This finding contrasts with accuracy-oriented multi-agent debate, where heterogeneity does not reduce convergence, suggesting that model diversity operates differently when there is no objectively correct answer. Second, coherence validation (using a frontier model to assess whether each evaluator's reasoning is genuinely grounded in its assigned values) reveals a \emph{fidelity--diversity tradeoff}: on a scenario with a dominant option, coherence validation further reduces concentration (46.1\% $\rightarrow$ 40.8\%, $p = 0.004$) by downweighting low-coherence majority voices, but on a scenario with genuinely competitive options, it increases concentration (22.9\% $\rightarrow$ 26.6\%, $p = 0.96$) by amplifying high-coherence evaluators who happen to cluster on one option. This tradeoff, where reasoning quality assessment and diversity preservation become competing objectives, may be a general property of multi-agent deliberation systems that employ any form of quality weighting, though confirmation across additional scenarios is needed. We report negative results from three failed Delphi designs, demonstrate that 8B models exhibit binary rather than graded responses to counter-arguments, and propose that the trustworthy tension rate (the fraction of value conflicts where both sides reason faithfully) provides a diagnostic measure of current small-model deliberation capabilities. Across both scenarios, roughly half of theoretical value conflicts are trustworthy, establishing a baseline for this model class.
\end{abstract}

\section{Introduction}

The use of LLM-based agents in multi-agent deliberation has attracted growing interest as a tool for policy analysis, stakeholder simulation, and participatory design \cite{chan2024chateval, du2023improving, liang2023encouraging}. In these systems, multiple agents are assigned different perspectives or roles and asked to evaluate policy options, with the goal of surfacing the value tensions inherent in complex decisions. If the system works, it should reveal which stakeholder perspectives favor which options and why; this information is valuable precisely because it preserves disagreement rather than resolving it.

In practice, however, these systems exhibit a persistent failure mode: artificial consensus. Evaluator agents converge on the same option regardless of their assigned value perspective, producing an illusion of agreement that obscures the genuine trade-offs underlying the policy question \cite{pitre2025consensagent, park2025social}. This failure is hard to detect because it mimics a desirable property, consensus, while actually representing a collapse of the deliberative process.

This failure mode is specific to \emph{normative} deliberation, where no objectively correct answer exists. In accuracy-oriented multi-agent debate, where agents collaborate to find a factual or mathematical answer, convergence is the goal, and recent work shows that debate forms a martingale on belief in the correct answer, providing no expected gain beyond voting \cite{huang2025debate}. Policy deliberation inverts this relationship: convergence is a failure mode, not a success criterion, because the system's value lies in mapping where different perspectives lead, not in reaching agreement.

We identify two distinct causes of artificial consensus in multi-agent LLM systems. The first is shared inductive bias: when all agents use the same underlying model, they share the same training data, RLHF preferences, and reasoning patterns. Even with different system prompts, the agents are fundamentally the same reasoner asked to wear different hats. The second is debate capture: agents assigned minority value perspectives are persuaded by the arguments of the majority during the debate phase, abandoning their assigned viewpoint in favor of the position that sounds most compelling. Recent work suggests this convergence reflects structured asymmetric persuasion dynamics rather than simple sycophantic conformity \cite{betz2025selective}, with format-dependent and model-specific patterns \cite{chen2025deliberative}.

This paper makes three contributions. First, we demonstrate that architectural heterogeneity, using a different LLM for each value perspective, significantly reduces artificial consensus in normative multi-agent deliberation, with large effect sizes across both tested scenarios. While recent work has established that heterogeneity improves reasoning accuracy \cite{chen2024reconcile} and produces richer behavioral differentiation \cite{elkandoussi2026whoami, maslowski2026hde}, its effect on normative deliberation---where convergence is a failure mode, not a success criterion---has not been tested. Our finding is domain-specific: heterogeneous model pools have been tested in accuracy-oriented debate without reducing convergence \cite{fang2025ahmad}, suggesting that model diversity operates through different mechanisms when there is no correct answer to converge on. Second, we introduce a coherence validation layer that uses a frontier model to assess whether each evaluator's reasoning is genuinely grounded in its assigned value perspective, and document a \emph{fidelity--diversity tradeoff}: quality assessment and diversity preservation can become competing objectives when model capability varies across value perspectives. Third, we report negative results from three Delphi designs and the binary response pattern of 8B models to counter-arguments, providing empirical constraints on multi-agent architecture design at this model scale.

The AI Council runs entirely on consumer hardware using seven locally-hosted 7--9B parameter models, with a single API call to a frontier model for coherence validation. We test it on two policy scenarios, child welfare intervention and urban housing policy, chosen to represent different structural properties (dominant option vs.\ genuinely competitive options).

\section{Related Work}

\textbf{Multi-agent debate and deliberation.} Multi-agent debate systems have been shown to improve reasoning quality on factual and mathematical tasks \cite{du2023improving, liang2023encouraging}. ChatEval \cite{chan2024chateval} uses multi-agent discussion for text evaluation, and several works explore society of mind architectures where LLM agents collaborate. However, these systems typically optimize for convergence on a correct answer, whereas our goal is the opposite: preserving legitimate disagreement when no objectively correct answer exists. Xiong et al.\ \cite{xiong2023examining} document conformity effects in LLM-based social simulations, where agents adopt majority positions, consistent with our observation of debate capture.

Huang et al.\ \cite{huang2025debate} prove that multi-agent debate forms a martingale on belief in the correct answer, providing no expected gain beyond independent voting for accuracy tasks. This theoretical result is complementary to our empirical finding: where their result establishes that debate adds nothing in accuracy contexts, our work shows that debate actively \emph{harms} in normative contexts through capture effects, motivating the architectural interventions we test.

\textbf{Sycophancy and convergence dynamics.} The tendency of RLHF-trained models to agree with users, known as sycophancy, is well-documented \cite{perez2023discovering, sharma2023towards}. Pitre et al.\ \cite{pitre2025consensagent} identify sycophancy as a core challenge in multi-agent debate specifically, finding that agents systematically converge toward shared positions. Park et al.\ \cite{park2025social} measure convergence scores exceeding 0.88 between opposing personas, quantifying the severity of artificial consensus in social simulation contexts.

However, the dynamics underlying convergence are more nuanced than simple conformity. Betz et al.\ \cite{betz2025selective} demonstrate that LLM opinion change follows structured patterns of asymmetric persuasion rather than blind sycophantic agreement. Chen et al.\ \cite{chen2025deliberative} find that consensus arises from a mix of inertia and conformity, with model-specific value patterns and format-dependent sycophancy effects. Our finding that 8B models exhibit binary (maintain or capitulate) rather than graded responses to counter-arguments extends this literature to the specific context of structured normative deliberation.

\textbf{Heterogeneous multi-agent systems.} The role of architectural heterogeneity in multi-agent LLM systems is an area of growing empirical investigation. Chen et al.\ \cite{chen2024reconcile} demonstrate in ReConcile that diverse model pools with confidence-weighted voting improve reasoning accuracy, establishing that model diversity is a productive design principle for collaborative problem-solving. Fang et al.\ \cite{fang2025ahmad} introduce A-HMAD, a heterogeneous multi-agent debate framework, and critically find that heterogeneity does not reduce convergence in accuracy-oriented tasks. Concurrent work extends these findings along two axes: Mas{\l}owski and Chudziak \cite{maslowski2026hde} show that architectural heterogeneity prevents argumentative degeneration in ethical tutoring dialogues, and El Kandoussi \cite{elkandoussi2026whoami} finds that heterogeneous agent groups exhibit significantly richer behavioral differentiation than homogeneous groups across multiple trait dimensions.

Our work applies this principle to a context with different success criteria: normative multi-stakeholder deliberation, where the failure mode is not reduced behavioral variety but artificial consensus on contested policy questions, and where coherence validation introduces a fidelity--diversity tradeoff absent from prior work. The divergence between Fang et al.'s null result for accuracy tasks and our large effects for normative tasks supports a domain-specific interpretation: architectural diversity disrupts shared inductive bias, which drives artificial consensus only when no objective ground truth constrains the outcome. Li et al.\ \cite{li2025talk} find that weak models in heterogeneous debate can degrade outcomes, a finding consistent with our observation that model quality variation across perspectives creates a fidelity--diversity tradeoff.

\textbf{Delphi method and structured expert judgment.} The Delphi method \cite{dalkey1963experimental, linstone1975delphi} is a structured process for eliciting and refining expert judgments through iterative rounds of anonymous feedback. Our coherence validation mechanism draws on Delphi principles, particularly the idea that expert responses should be assessed for the quality of their reasoning, not just their conclusions, but departs from traditional Delphi in a critical way: we do not aim for convergence. We use the term first-choice concentration throughout this paper to avoid the ambiguity of ``convergence,'' which is desirable in traditional Delphi but represents failure in our framework.

\textbf{AI-assisted policy simulation.} Recent work has explored using LLMs for policy analysis \cite{argyle2023out}, stakeholder modeling, and public consultation. Sorek et al.\ \cite{sorek2024child} study child welfare intervention policy in Israel, providing the empirical grounding for one of our test scenarios. Our work is distinguished by its focus on the structural conditions under which multi-agent deliberation produces reliable outputs for normative questions, rather than on the specific policy recommendations generated.

\section{System Architecture}

The AI Council is a three-phase deliberation system designed to simulate multi-stakeholder policy deliberation while preserving value-based disagreement. Each deliberation proceeds through structured debate, independent evaluation, and (optionally) coherence validation.

\subsection{Phase 1: Structured Debate}

Three champion agents each advocate for one of three policy options across three rounds:

\begin{enumerate}
    \item \textbf{Position papers:} Each champion presents the case for its assigned option.
    \item \textbf{Critiques:} Each champion critiques the other two options from its own value perspective---for example, the Security-focused champion critiques alternatives on safety grounds, not from a neutral analytical stance.
    \item \textbf{Defenses:} Each champion responds to critiques of its assigned option.
\end{enumerate}

Champions are assigned fixed personality--option pairings across all runs: the Conservative (Security Focus) personality advocates for Option 1, the Innovator (Risk Tolerance) for Option 2, and the Pragmatist (Pragmatism) for Option 3. In heterogeneous configurations, each champion uses its assigned model; in the homogeneous baseline, all champions use the same base model.

The debate phase produces a structured argument corpus that is provided in full to all evaluators. Because each champion critiques competing options through its own value lens, the corpus is itself shaped by value perspectives: each option's weaknesses are surfaced by the value dimensions most likely to find them problematic. Champions do not see each other's output during the same round.

\subsection{Phase 2: Independent Evaluation}

Seven evaluator agents, each primed with a distinct value perspective, independently read the complete debate output and produce a ranking of all three options with supporting reasoning. Evaluators do not see each other's output at any point.

Each evaluator receives a system prompt specifying its value perspective name, a natural-language definition of that perspective, and instructions to rank options based on alignment with the perspective. Evaluators produce a first choice, second choice, and third choice, along with a reasoning paragraph.

\subsection{Value Perspectives}

We define six value dimensions, each representing a coherent orientation toward policy evaluation:

\begin{itemize}
    \item \textbf{Security Focus:} Prioritizes safety, stability, proven methods, and risk mitigation.
    \item \textbf{Risk Tolerance:} Favors bold approaches, accepts uncertainty, and challenges the status quo.
    \item \textbf{Pragmatism:} Emphasizes practical feasibility, real-world constraints, and implementation.
    \item \textbf{Performance Focus:} Values efficiency, speed, measurable results, and return on investment.
    \item \textbf{Simplicity Preference:} Prefers straightforward, uncomplicated, easy-to-implement solutions.
    \item \textbf{Creativity:} Seeks novel, context-specific, and innovative solutions.
\end{itemize}

In the current implementation, the Innovator role's tied trait parameters (Risk Tolerance and Creativity both at 0.9) resolve to Risk Tolerance as the primary perspective, due to Python dictionary key ordering in the trait resolution function. This leaves Creativity without a dedicated evaluator. The system therefore tests five dedicated value perspectives, with Creativity partially represented through the Innovator's secondary trait priming. Tension pairs involving Creativity (Security--Creativity, Pragmatism--Creativity, Simplicity--Creativity) are tested indirectly through the Risk Tolerance evaluator rather than through an evaluator whose sole orientation is creative novelty. Future implementations should adjust trait parameters to ensure all six perspectives are independently represented.

These dimensions generate a theoretical value tension map: pairs of perspectives that should, in principle, disagree about policy options. Six tension pairs are identified: Security--Risk Tolerance, Security--Creativity, Performance--Simplicity, Pragmatism--Creativity, Pragmatism--Risk Tolerance, and Simplicity--Creativity. The fraction of these pairs that actually disagree in practice provides a measure of the system's ability to preserve distinct viewpoints.

\subsection{Model Assignment}

Seven locally-hosted models (7--9B parameters) running via Ollama on consumer hardware are each paired with a specific evaluator role:

\begin{table}[t]
\centering
\small
\resizebox{\columnwidth}{!}{%
\begin{tabular}{lll}
\toprule
Evaluator Role & Value Perspective & Model \\
\midrule
Conservative & Security Focus & Qwen3-8B \\
Innovator & Risk Tolerance$^\dagger$ & Mistral-NeMo \\
Pragmatist & Pragmatism & Mistral-7B-Instruct-v0.3 \\
Perfectionist & Security Focus & Qwen2.5-Coder-7B \\
Minimalist & Simplicity Preference & Dolphin3-8B \\
Driver & Performance Focus & DeepSeek-R1-8B \\
Guardian & Security Focus & Gemma2-9B \\
\bottomrule
\end{tabular}}
\caption[Fixed model--role assignments for the heterogeneous configuration.]{Fixed model--role assignments for the heterogeneous configuration. Note that three evaluators share the Security Focus perspective but use different models, testing whether architectural diversity produces behavioral diversity even within the same value framing.\protect\footnotemark\ $^\dagger$The Innovator's tied trait parameters (Risk Tolerance and Creativity both at 0.9) resolve to Risk Tolerance as the primary perspective; Creativity is represented as a secondary trait only.}
\label{tab:models}
\end{table}

\footnotetext{The Perfectionist role's strongest trait parameter is Security Focus (0.8), not Performance Focus as the role name might suggest, resulting in three evaluators sharing this perspective despite different role names. Value perspective assignments are determined by each role's highest-weighted trait, not by role name.}

Models were selected for architectural diversity: the pool includes models from five different organizations spanning different training methodologies (instruction-tuned, code-specialized, reasoning-optimized, uncensored).

Model--role pairings were determined through an empirical profiling battery conducted before data collection. Each model was tested across all value perspectives on standardized policy prompts, with alignment assessed through keyword-frequency scoring: for each perspective, a set of perspective-characteristic keywords was defined (e.g., ``safety,'' ``stability,'' ``proven'' for Security Focus; ``efficiency,'' ``measurable,'' ``results'' for Performance Focus), and each model's responses were scored by the frequency of perspective-aligned keywords normalized by response length. The model with the highest alignment score for each role was assigned to it (e.g., Dolphin3-8B scored 0.571 on simplicity alignment, the highest of any model tested; Mistral-NeMo scored 0.429 on creativity alignment, the highest for that dimension). Models showed differentiated value profiles rather than uniform quality: Qwen3-8B scored 0.571 on security but 0.000 on pragmatism; Dolphin3-8B scored 0.571 on simplicity but 0.125 on performance. Full profiling methodology, keyword lists, and per-model results are available in supplementary materials.

This configuration results in three evaluators sharing the Security Focus perspective (Conservative, Perfectionist, Guardian), each using a different model. While this reduces the number of distinct perspectives tested, it provides a natural test of within-perspective model diversity: do different models produce different behavior when assigned the same value framing?

Across the child welfare scenario, the three Security Focus evaluators showed substantial behavioral divergence---the Guardian (Gemma2-9B, mean coherence $\approx$0.97) consistently voted for established methods, while the Perfectionist (Qwen2.5-Coder-7B, mean coherence $\approx$0.59) frequently drifted toward performance-based reasoning. This divergence under identical value priming provides additional evidence that architectural heterogeneity, not just value diversity, drives behavioral differences.

\section{Coherence Validation (Delphi)}

\subsection{Design}

The coherence validation layer is a post-processing step applied after evaluation is complete. It does not modify the deliberation itself. A single call to a frontier model (Claude Sonnet 4, Anthropic 2025) assesses each evaluator's reasoning for coherence with their assigned value perspective.

Critically, the frontier model receives only:

\begin{itemize}
    \item The evaluator's assigned value perspective name (e.g., ``Security Focus'')
    \item The natural-language definition of that perspective
    \item The evaluator's reasoning text
\end{itemize}

The frontier model does not receive: the evaluator's votes or rankings, the names of the policy options, any other evaluator's output, or the debate transcript. This information restriction ensures that coherence scoring reflects reasoning quality rather than agreement with any particular position.

We acknowledge two potential confounds in this design. First, evaluator reasoning may reference policy options by name, allowing the frontier model to potentially infer which option was chosen. Since coherence scoring assesses the relationship between reasoning and assigned value perspective---not between reasoning and a ``correct'' vote---knowledge of the chosen option does not provide a direct basis for biased scoring. Second, and more subtly, some option--perspective pairings may produce reasoning that is inherently easier to score as coherent because the vocabulary maps naturally: a Security Focus evaluator supporting an intensive family preservation program will use words like ``stability'' and ``proven'' that are literally the perspective's characteristic vocabulary, while the same evaluator supporting deregulation must work harder to connect its reasoning to security values. This means coherence scores may partially reflect how naturally an option maps onto a perspective's vocabulary, not just the quality of the evaluator's reasoning. We cannot fully disentangle this confound from genuine coherence differences without a fundamentally different validation design, and note it as a limitation.

The frontier model returns a coherence score in $[0.0, 1.0]$ for each evaluator, representing the degree to which the evaluator's reasoning is genuinely grounded in the assigned value perspective. We note that coherence scores reflect the frontier model's judgment about what constitutes faithful value-based reasoning, not an objective ground truth. The test-retest reliability (Section~\ref{sec:reliability}) establishes that this judgment is consistent, and cross-model validation (Section~\ref{sec:crossmodel}) confirms that a different frontier model produces strongly correlated scores, though absolute levels may differ. Relative differences between evaluators, specifically which perspectives are scored higher or lower, are more interpretable than absolute coherence levels.

\subsection{Score Application}

Coherence scores are used as weights in a modified Borda count. In the unweighted scheme (State B), each evaluator contributes equally: their first choice receives 2 points, second choice 1 point, third choice 0 points. In the weighted scheme (State C), each evaluator's Borda points are multiplied by their coherence score. A low-coherence evaluator still contributes, but proportionally less.

This design preserves all votes (no evaluator is silenced) while reducing the influence of evaluators whose reasoning does not reflect their assigned perspective. The Voice Authenticity Rate (fraction of evaluators scoring $\geq 0.6$) provides a summary measure of simulation quality.

\subsection{Journey Through Failed Designs}

The current coherence validation mechanism (v8) was preceded by three failed designs that are informative about the constraints on Delphi-style interventions with small language models.

\textbf{Delphi v6.0 (peer exposure).} Evaluators were shown summaries of all other evaluators' critiques and asked to reconsider their positions. Models were overwhelmed by multiple persuasive critiques and fled to whichever option had not been criticized, typically a third option that no evaluator had originally endorsed. Peer exposure amplified rather than corrected artificial consensus.

\textbf{Delphi v6.1 (value-tension filtered).} To address the overload problem, evaluators were shown only critiques from their designated tension partners (e.g., a Security Focus evaluator sees only the Risk Tolerance critique). Even a single relevant counter-argument destabilized 8B models. Two rounds of this exposure compounded sycophantic flips. \textbf{Delphi v7 (self-confrontation).} Evaluators were asked to re-read their own reasoning and identify whether it was consistent with their assigned value perspective. Across 10 test runs, zero vote changes occurred; the models re-read their own reasoning and simply reiterated it, demonstrating that post-hoc rationalization exceeds self-reflection capability at this model scale.

These failures established an empirical constraint: 8B models have a binary response to counter-arguments: they either maintain their position entirely or capitulate entirely. No ``consider and reject'' middle state exists. This constraint motivated the move to an external validation architecture where the deliberation itself is not modified.

\subsection{Test-Retest Reliability}
\label{sec:reliability}

To assess the stability of coherence scoring, 10 deliberations (70 evaluator assessments) were scored twice under identical conditions. Reliability was good: ICC(3,1) = 0.849, Pearson $r = 0.876$, Spearman $\rho = 0.861$. The mean absolute difference between test and retest scores was 0.124, and 86\% of scores were stable within $\pm 0.2$. A slight systematic upward bias on retest was observed (mean shift +0.07), suggesting that initial coherence scores may be marginally conservative. High-coherence evaluators (Guardian, Driver) showed the greatest stability; mid-range evaluators (Perfectionist, Minimalist) showed more variation.

\subsection{Cross-Model Robustness}
\label{sec:crossmodel}

To test whether coherence scores reflect properties of the reasoning text rather than artifacts of a single model's preferences, we ran a second frontier model (OpenAI GPT-5.2) as an independent coherence judge on a randomly sampled subset of 10 deliberations per scenario (140 evaluator assessments total), using the identical prompt and calibration anchors. Pearson correlations between GPT-5.2 and Claude Sonnet scores were $r = 0.74$ (child welfare) and $r = 0.88$ (housing), both $p < 0.001$; Spearman rank correlations were $\rho = 0.77$ and $\rho = 0.89$ respectively. Claude assigned slightly higher scores on average (mean difference ${\sim}0.09$), but the rank ordering was preserved: both models identified the same evaluators as strongly or weakly coherent, and the relative ordering across perspectives (Performance Focus highest, Pragmatism and Simplicity Preference lowest) was consistent across judges.

\section{Experimental Design}

\subsection{Three-State Design}

We employ a three-state experimental design:

\begin{itemize}
    \item \textbf{State A (Homogeneous):} All 7 evaluators and 3 champions use the same base model (Mistral-7B-Instruct-v0.3). Value priming is applied through system prompts, but all agents share the same underlying architecture. This tests single-model convergence.
    \item \textbf{State B (Heterogeneous):} Each evaluator uses a different model matched to its value perspective (Table~\ref{tab:models}). Champions also use their assigned models. No coherence validation is applied. This tests the effect of architectural diversity alone.
    \item \textbf{State C (Heterogeneous+Delphi):} The same deliberations as State B, post-processed through coherence validation. States B and C are paired; they derive from the same underlying runs, with C applying coherence-weighted scoring to B's outputs.
\end{itemize}

State A uses 25 runs per scenario. State B uses 35 runs per scenario, which also generates 35 paired State C results. The total is 60 deliberations per scenario and 120 overall.

\subsection{Scenarios}

Two scenarios were selected to test the system under structurally different conditions.

\textbf{Scenario 1: Child Welfare Intervention.} Derived from research on Israeli child welfare policy \cite{sorek2024child}, this scenario asks: How should child welfare services balance family preservation with child safety? Three options are presented: (A) an intensive family preservation program with small caseloads and frequent home visits; (B) out-of-home placement with permanency planning; (C) standard welfare department practice with large caseloads. Option A is inherently more appealing (it offers more resources per family), creating a known asymmetry. This scenario tests the system's behavior when one option is dominant.

\textbf{Scenario 2: Urban Housing Crisis.} This scenario asks: What policy approach should cities adopt to address the housing affordability crisis? Three options are presented: (A) strict rent control with tenant protections; (B) deregulation of zoning with construction incentives; (C) public housing expansion funded by new taxes. These options represent fundamentally different policy philosophies with no inherent quality hierarchy, testing the system's behavior when genuine three-way competition exists.

\subsection{Metrics}

\textbf{First-Choice Concentration.} The normalized agreement on the top-choice option:
\begin{equation}
\text{FCC} = \frac{\max(v_1, v_2, v_3) - \lfloor N/K \rfloor}{N - \lfloor N/K \rfloor}
\end{equation}
where $v_i$ is the number of first-choice votes for option $i$, $N = 7$ evaluators, and $K = 3$ options. With these parameters, FCC takes exactly five discrete values: 14.3\%, 35.7\%, 57.1\%, 78.6\%, and 100\%.

We note an important caveat: the current system assigns three of seven evaluators to the Security Focus perspective (Section~\ref{tab:models}). This means FCC and vote distributions are structurally weighted toward Security Focus preferences; the effective number of independent perspective-bearing evaluators is five rather than seven. The A vs.\ B heterogeneity comparison remains valid because the same 3/7 structure is present in both states, but absolute FCC values should be interpreted as reflecting this perspective imbalance. Future implementations with seven distinct perspectives would produce more balanced vote distributions.

\textbf{Borda Margin.} The gap between the first- and second-ranked options using the full Borda count (first choice = 2 points, second = 1, third = 0), normalized by the maximum possible score:
\begin{equation}
\text{BM} = \frac{S_1 - S_2}{2N}
\end{equation}
where $S_1$ and $S_2$ are the Borda scores of the first- and second-ranked options. This captures ranking depth, not just top-choice agreement.

\textbf{Effective Perspectives.} The Shannon entropy of the first-choice vote distribution:
\begin{equation}
H = -\sum_{i=1}^{K} p_i \log_2(p_i)
\end{equation}
where $p_i$ is the fraction of first-choice votes for option $i$. Higher entropy indicates more distinct viewpoints represented.

\textbf{Voice Authenticity Rate.} The fraction of evaluators with coherence score $\geq 0.6$. This is a simulation quality metric applicable only to State C.

\textbf{Trustworthy Tension Rate.} Of the theoretical tension pairs present in a deliberation, the fraction where the quality of disagreement (or agreement) can be trusted, meaning either both sides are coherently representing their values and disagree (authentic disagreement), or both sides are coherent and happen to agree (genuine agreement). The complementary categories are suspect agreement (both sides agree but at least one has low coherence, suggesting debate capture) and partial (one side has low coherence, making the interaction uninterpretable).

\subsection{Statistical Approach}

States A vs.\ B are compared using the Mann-Whitney U test (one-tailed: A $>$ B for concentration, A $<$ B for perspectives), as these are independent samples. States B vs.\ C are compared using the Wilcoxon signed-rank test (one-tailed), as these are paired observations. The primary test (concentration for A vs.\ B; Borda margin for B vs.\ C) uses $\alpha = 0.05$. Secondary tests use Bonferroni-corrected $\alpha = 0.0125$. Effect sizes are reported as rank-biserial $r$, with 95\% confidence intervals from 10,000-resample bootstrapping.

We note that the discrete structure of FCC (five possible values; see Appendix~A) means that statistical tests on this metric should be interpreted as comparisons of distributional tendency over repeated runs rather than as measurements of a continuous effect. The bootstrapped confidence intervals and the use of Borda margin as a complementary continuous metric mitigate this concern, but we advise against over-interpreting small FCC differences. In effect, statistical tests on FCC compare the mixture probabilities across the five discrete values rather than measuring a continuous effect.

\section{Results}

\subsection{Scenario 1: Child Welfare Intervention}

Table~\ref{tab:cw} summarizes the primary metrics across all three states for the child welfare scenario.

\begin{table*}[t]
\centering
\small
\begin{tabular}{lcccc}
\toprule
Metric & State A & State B & State C & 95\% CI \\
 & Homogeneous & Heterogeneous & Het.+Delphi & \\
\midrule
$N$ (runs) & 25 & 35 & 35 (paired) & \\
FCC (mean $\pm$ SD) & $.709 \pm .194$ & $.461 \pm .223$ & $.408 \pm .228$ & A--B: $[.140, .349]$ \\
Borda Margin (mean $\pm$ SD) & $.326 \pm .067$ & $.218 \pm .130$ & $.184 \pm .102$ & B--C: $[.011, .059]$ \\
Eff.\ Perspectives (mean) & 0.74 & 1.07 & 1.12 & \\
Voice Authenticity & --- & --- & 80.4\% & \\
Mean Coherence & --- & --- & 0.76 & \\
\midrule
\multicolumn{5}{l}{\textit{Winner distribution}} \\
Intensive & 25/25 & 32/35 & 29/35 & \\
Out-of-home & 0 & 1 & 2 & \\
Standard & 0 & 1 & 2 & \\
Tie & 0 & 1 & 2 & \\
\bottomrule
\end{tabular}
\caption{Child welfare scenario: summary statistics across states. State C applies coherence-weighted Borda scoring to the same deliberations as State B. FCC = first-choice concentration. The 95\% CI column shows bootstrap confidence intervals for the primary comparison at each stage: A vs.\ B for FCC (Mann-Whitney), B vs.\ C for Borda margin (Wilcoxon).}
\label{tab:cw}
\end{table*}

\textbf{Heterogeneity effect (A vs.\ B).} Architectural heterogeneity significantly reduced first-choice concentration from 70.9\% to 46.1\% (Mann-Whitney $U = 689.5$, $p < 0.001$, $r = 0.58$, large effect). Effective perspectives increased from 0.74 to 1.07 ($p = 0.003$, $r = 0.43$, medium effect; significant after Bonferroni correction).

\textbf{Coherence validation effect (B vs.\ C).} Coherence weighting further reduced Borda margin from 0.218 to 0.184 (Wilcoxon $W = 457$, $p = 0.010$, $r = 0.45$, medium effect). First-choice concentration dropped from 46.1\% to 40.8\% ($p = 0.004$, $r = 0.52$, large effect; Bonferroni-significant). Effective perspectives increased from 1.07 to 1.12 ($p < 0.001$, $r = 0.61$, large effect; Bonferroni-significant). Margins were reduced in 22 of 35 runs and increased in 12.

\textbf{Winner diversity.} Despite these improvements, the Intensive family preservation option won 25/25 State A runs and 32/35 State B runs. The system detects the dominance of this option but does not create disagreement where the underlying models agree---heterogeneity introduces minority voices, but the majority position remains strong.

\textbf{Archetype stability.} Table~\ref{tab:cw_arch} shows how each value perspective behaved across State B runs, explaining why coherence validation reduces concentration in this scenario.

\begin{table}[t]
\centering
\small
\resizebox{\columnwidth}{!}{%
\begin{tabular}{lccc}
\toprule
Perspective & Most Common & Consist. & Mean Coh. \\
\midrule
Performance Focus & Intensive & 49\% & 0.91 \\
Pragmatism & Intensive & 94\% & 0.59 \\
Risk Tolerance & Intensive & 91\% & 0.79 \\
Security Focus & Intensive & 49\% & 0.81 \\
Simplicity Pref. & Intensive & 57\% & 0.61 \\
\bottomrule
\end{tabular}}
\caption{Child welfare scenario: archetype stability across State B runs. Nearly all perspectives favor Intensive, but low-coherence evaluators (Pragmatism, Simplicity) who agree with the majority are downweighted, amplifying the minority voices that do dissent.}
\label{tab:cw_arch}
\end{table}

All perspectives converge on the Intensive option, but with varying coherence. Low-coherence evaluators who agree with the majority (Pragmatism at 0.59, Simplicity at 0.61) are downweighted by coherence validation, which has the net effect of slightly amplifying the influence of minority votes from higher-coherence dissenters. Coherence validation reduces concentration in this scenario not by creating disagreement but by amplifying what little disagreement exists.

\subsection{Scenario 2: Urban Housing Crisis}

Table~\ref{tab:housing} presents results for the housing scenario, which displays markedly different behavior.

\begin{table*}[t]
\centering
\small
\begin{tabular}{lcccc}
\toprule
Metric & State A & State B & State C & 95\% CI \\
 & Homogeneous & Heterogeneous & Het.+Delphi & \\
\midrule
$N$ (runs) & 25 & 35 & 35 (paired) & \\
FCC (mean $\pm$ SD) & $.460 \pm .263$ & $.229 \pm .119$ & $.266 \pm .148$ & A--B: $[.120, .342]$ \\
Borda Margin (mean $\pm$ SD) & $.238 \pm .145$ & $.099 \pm .085$ & $.127 \pm .107$ & B--C: $[-.054, -.002]$ \\
Eff.\ Perspectives (mean) & 1.04 & 1.43 & 1.36 & \\
Voice Authenticity & --- & --- & 75.9\% & \\
Mean Coherence & --- & --- & 0.75 & \\
\midrule
\multicolumn{5}{l}{\textit{Winner distribution (first-choice)}} \\
Rent Control & 0 & 15 & 15 & \\
Deregulate & 1 & 3 & 3 & \\
Public Housing & 23 & 12 & 12 & \\
Tie & 1 & 5 & 5 & \\
\bottomrule
\end{tabular}
\caption{Housing scenario: summary statistics across states. State C applies coherence-weighted Borda scoring to the same deliberations as State B. First-choice vote distributions are identical between B and C (weighting affects Borda scores, not votes); however, the weighted Borda winner differs from the unweighted winner in 19 of 35 runs. FCC = first-choice concentration. The 95\% CI column shows bootstrap confidence intervals for the primary comparison at each stage. Note that the B--C Borda margin CI is entirely negative, indicating that coherence weighting increases margins in this scenario.}
\label{tab:housing}
\end{table*}

\textbf{Heterogeneity effect (A vs.\ B).} Architectural heterogeneity halved first-choice concentration from 46.0\% to 22.9\% (Mann-Whitney $U = 656.5$, $p < 0.001$, $r = 0.50$, large effect). Effective perspectives rose from 1.04 to 1.43 ($p < 0.001$, $r = 0.61$, large effect; Bonferroni-significant). Crucially, winner diversity also improved dramatically: State A was dominated by Public Housing (23/25), whereas State B produced three-way competition (Rent Control 15, Public Housing 12, Deregulate 3, Tie 5).

\textbf{Coherence validation effect (B vs.\ C): reversed direction.} On the housing scenario, coherence validation moved all metrics in the opposite direction from child welfare. Borda margin increased from 0.099 to 0.127 (Wilcoxon $p = 0.96$, $r = -0.34$; the one-tailed test for margin reduction is non-significant because the effect runs the other way). First-choice concentration increased from 22.9\% to 26.6\%. Effective perspectives decreased from 1.43 to 1.36. Margins increased in 23 of 35 runs and decreased in 12.

This reversal is the central finding regarding coherence validation and warrants detailed examination.

\textbf{Explaining the reversal.} The reversal is explained by the distribution of coherence scores across value perspectives (Table~\ref{tab:housing_arch}).

\begin{table}[t]
\centering
\small
\resizebox{\columnwidth}{!}{%
\begin{tabular}{lccc}
\toprule
Perspective & Most Common & Consist. & Mean Coh. \\
\midrule
Performance Focus & Deregulate & 74\% & 0.95 \\
Risk Tolerance & Deregulate & 74\% & 0.86 \\
Security Focus & Rent Control & 67\% & 0.72 \\
Pragmatism & Public Housing & 91\% & 0.63 \\
Simplicity Pref. & Rent Control & 49\% & 0.61 \\
\bottomrule
\end{tabular}}
\caption{Housing scenario: archetype stability across State B runs. High-coherence perspectives (Performance Focus, Risk Tolerance) cluster on Deregulate; low-coherence perspectives are spread across Rent Control and Public Housing.}
\label{tab:housing_arch}
\end{table}

The two highest-coherence perspectives, Performance Focus (0.95) and Risk Tolerance (0.86), both cluster on the Deregulate option. The two lowest-coherence perspectives, Pragmatism (0.63) and Simplicity Preference (0.61), are spread across Rent Control and Public Housing. When coherence weighting is applied, the Deregulate coalition is amplified relative to the dispersed low-coherence voices, increasing concentration.

The coherence validation layer is working correctly in both scenarios: it identifies which evaluators reason most faithfully from their assigned values. But fidelity and diversity trade off when model quality varies across value perspectives. The models assigned to Performance Focus (DeepSeek-R1-8B) and Risk Tolerance (Mistral-NeMo) happen to produce more coherent reasoning than those assigned to Pragmatism (Mistral-7B) and Simplicity Preference (Dolphin3-8B). This is a property of the model pool, not the validation mechanism.

\subsection{Cross-Scenario Comparison}

\textbf{Tension quality.} The Trustworthy Tension Rate provides a summary diagnostic of deliberation quality by examining each theoretical tension pair (Table~\ref{tab:tension}).

\begin{table}[t]
\centering
\small
\resizebox{\columnwidth}{!}{%
\begin{tabular}{lcc}
\toprule
Tension Category & Child Welfare & Housing \\
\midrule
Authentic disagreement & 34.3\% & 34.3\% \\
Genuine agreement & 21.0\% & 10.5\% \\
\textbf{Trustworthy (total)} & \textbf{55.2\%} & \textbf{44.8\%} \\
Suspect agreement & 21.0\% & 7.6\% \\
Partial (one side low-coh.) & 23.8\% & 47.6\% \\
\bottomrule
\end{tabular}}
\caption{Tension quality distribution across scenarios. Trustworthy tensions are those where both sides are coherently representing their assigned values, whether they disagree (authentic) or agree (genuine). Note: three of six tension pairs involve Creativity, which lacks a dedicated evaluator (Section~3.3) and is tested only indirectly through the Risk Tolerance evaluator. Rates for the three direct tension pairs (Security--Risk Tolerance, Performance--Simplicity, Pragmatism--Risk Tolerance) versus the three indirect pairs may differ; we report the aggregate here.}
\label{tab:tension}
\end{table}

Across both scenarios, approximately half of theoretical value tensions are trustworthy (55.2\% child welfare, 44.8\% housing). The housing scenario has a higher rate of partial tensions (47.6\% vs.\ 23.8\%), consistent with its lower overall coherence and the greater variation in model quality across perspectives. This rate is best understood as a measurement of current small-model deliberation capabilities rather than an architectural limitation.

\textbf{Statistical summary.} Table~\ref{tab:stats} consolidates all statistical tests across both scenarios.

\begin{table*}[t]
\centering
\small
\begin{tabular}{llllccc}
\toprule
Scenario & Test & Comparison & Metric & $p$ & $r$ & Sig.? \\
\midrule
CW & Mann-Whitney & A vs B & FCC & $<.001$ & 0.58 (L) & Yes \\
CW & Mann-Whitney & A vs B & Perspectives & .003 & 0.43 (M) & Yes$^\dagger$ \\
CW & Wilcoxon & B vs C & Margin & .010 & 0.45 (M) & Yes \\
CW & Wilcoxon & B vs C & FCC & .004 & 0.52 (L) & Yes$^\dagger$ \\
CW & Wilcoxon & B vs C & Perspectives & $<.001$ & 0.61 (L) & Yes$^\dagger$ \\
Housing & Mann-Whitney & A vs B & FCC & $<.001$ & 0.50 (L) & Yes \\
Housing & Mann-Whitney & A vs B & Perspectives & $<.001$ & 0.61 (L) & Yes$^\dagger$ \\
Housing & Wilcoxon & B vs C & Margin & .959 & $-$0.34 (M) & No$^*$ \\
Housing & Wilcoxon & B vs C & FCC & .959 & $-$0.34 (M) & No$^*$ \\
Housing & Wilcoxon & B vs C & Perspectives & .999 & 0.72 (L) & No$^*$ \\
\bottomrule
\end{tabular}
\caption{Statistical tests across both scenarios. L = large, M = medium effect size. $^\dagger$Significant after Bonferroni correction ($\alpha = 0.0125$). $^*$Effect in opposite direction from hypothesized; one-tailed $p$-values near 1.0 reflect that the data support the reverse hypothesis.}
\label{tab:stats}
\end{table*}

The heterogeneity effect (A vs.\ B) is robust: large, significant effects in both scenarios across all metrics. The coherence validation effect (B vs.\ C) is scenario-dependent: significant reductions in all child welfare metrics, but reversed (non-significant) increases in all housing metrics. This divergence is not a failure of the system but a genuine finding about the conditions under which fidelity assessment and diversity preservation align or conflict.

\section{Discussion}

\subsection{Architectural Heterogeneity in Normative Deliberation}

The most robust finding of this study is that architectural heterogeneity, using a different model for each value perspective, significantly reduces artificial consensus in normative multi-agent deliberation. The effect is large and consistent across both scenarios, with first-choice concentration dropping by 25 percentage points (child welfare) and 23 percentage points (housing) when moving from homogeneous to heterogeneous configurations. This finding converges with concurrent work showing that heterogeneity produces richer behavioral differentiation \cite{elkandoussi2026whoami} and prevents argumentative degeneration \cite{maslowski2026hde}, while extending these results to the specific context of normative policy deliberation where the failure mode is artificial consensus rather than reasoning inaccuracy or dialectical stagnation.

This result must be interpreted in the context of the accuracy-oriented literature. Fang et al.\ \cite{fang2025ahmad} find that heterogeneous model pools do \emph{not} reduce convergence in accuracy-focused multi-agent debate, where the goal is to reach the correct answer. Our opposite finding in normative deliberation---where no correct answer exists---supports a domain-specific interpretation: architectural heterogeneity disrupts \emph{shared inductive bias}, which is a primary driver of artificial consensus only when convergence is not constrained by an objective ground truth. In accuracy tasks, models converge because the evidence points to the correct answer regardless of architectural variation; in normative tasks, they converge because they share training biases that favor certain value positions, and using different models breaks this spurious agreement.

In practice, researchers and practitioners building multi-agent deliberation systems for normative questions should use heterogeneous model pools as a default. The effect is large enough to be worth the engineering cost.

\subsection{The Fidelity--Diversity Tradeoff}

The divergent effects of coherence validation across scenarios reveal a tension in multi-agent deliberation design. When a coherence validation layer identifies that some evaluators reason more faithfully than others, applying this information necessarily alters the balance of voices. If the high-fidelity evaluators happen to agree with each other, fidelity assessment reduces diversity. If they happen to disagree, it increases diversity.

In the child welfare scenario, coherence validation reduces concentration because it downweights low-coherence evaluators who happen to agree with the majority (consensus captured). In the housing scenario, it increases concentration because the highest-coherence evaluators cluster on one option (Deregulate), while the voices providing diversity (Pragmatism, Simplicity Preference) happen to be modeled by weaker reasoners.

This is not a flaw in the coherence validation mechanism---the mechanism correctly identifies which evaluators reason most faithfully from their assigned values. Rather, it reflects a deeper structural fact: \emph{when model quality varies across value perspectives, fidelity and diversity are competing objectives}. Any system that assesses reasoning quality will amplify perspectives modeled by stronger reasoners, which may or may not align with the goal of preserving disagreement.

We propose that the appropriate response is not to abandon coherence validation but to report both weighted and unweighted results, treating the gap between them as informative. When B and C diverge, the divergence itself tells the analyst something: specific perspectives are being modeled less faithfully, and the disagreement surface should be interpreted with that caveat.

\subsection{Model Quality and Perspective--Model Fit}

The fidelity--diversity tradeoff raises the question of whether coherence differences reflect overall model quality or perspective--model fit. The profiling data provide partial evidence. Dolphin3-8B scored 0.571 on simplicity alignment in profiling---the highest of any model for that perspective---but achieved only 0.61 mean coherence in actual deliberations. This suggests a model well-matched to its perspective that nevertheless produces lower-quality reasoning under the pressure of multi-round debate. Conversely, DeepSeek-R1-8B (Performance Focus) achieves 0.95 mean coherence, consistent with its reasoning-optimized training. The pattern is consistent with findings from Li et al.\ \cite{li2025talk} that weaker models in heterogeneous debate degrade outcomes: model capability appears to interact with architectural diversity, such that the benefits of heterogeneity are partially offset when some models cannot maintain coherent value-based reasoning.

This confound---coherence scores as a joint measure of model capability and perspective--model fit---cannot be fully resolved without a combinatorial design testing each model across all perspectives. We note, however, that the differentiated profiling results (models scoring high on some perspectives and low on others) suggest that perspective--model fit is a meaningful component, not merely a proxy for overall quality.

To assess whether profiling alignment and deliberation coherence reflect the same underlying construct, we computed the correlation between each perspective's keyword-frequency profiling score and its mean coherence score across deliberations. The correlation is negative in both scenarios (child welfare: Pearson $r = -0.83$, $p = 0.079$; housing: $r = -0.94$, $p = 0.019$; $N = 5$ perspectives). Models that naturally echo a perspective's characteristic vocabulary do not necessarily produce higher-quality reasoned arguments from that perspective under multi-round debate pressure---the model with the highest profiling alignment (Mistral-7B, pragmatism: 0.667) produced the lowest mean coherence (0.59--0.63), while the model with the lowest profiling alignment (DeepSeek-R1-8B, performance: 0.375) produced the highest (0.91--0.95). These correlations are suggestive given the small sample, but they indicate that coherence validation is not simply recapitulating the profiling signal through a different measurement instrument.

\subsection{Binary Response Pattern in Small Models}

The failure of Delphi designs v6.0, v6.1, and v7 is consistent with a growing body of evidence on sycophancy in language models \cite{pitre2025consensagent, sharma2023towards}. Our specific contribution is the observation that 8B models exhibit binary rather than graded responses to counter-arguments: they either maintain their position entirely (including when asked to self-reflect) or capitulate entirely (when exposed to opposing arguments). The absence of a ``consider and reject'' middle state is not a universal property of language models---frontier models can engage in nuanced weighing of arguments---but it appears to be a robust characteristic of the 7--9B parameter range.

This finding has implications beyond our system. Any multi-agent architecture that exposes small-model agents to arguments from other agents risks inducing capitulation cascades, where the first agent to encounter a persuasive argument flips, creating a feedback loop. Architectural designs that preserve agent isolation during evaluation---as our system does---are necessary safeguards at this model scale.

\subsection{Scenario Structure Matters}

The contrast between our two scenarios illustrates that the effectiveness of a multi-agent deliberation system depends on the structural properties of the policy question being asked. Child welfare, with its dominant option (Intensive), reveals the system's ceiling: heterogeneity introduces minority perspectives but cannot overcome strong agreement when one option genuinely provides more resources. Housing, with its three-way competition, reveals the system's potential: heterogeneous evaluators distribute across options in patterns that reflect their assigned values.

An additional intervention---rewriting child welfare option descriptions to be more balanced---produced no meaningful change in convergence (48.6\% vs.\ 46--49\% baseline), confirming that debate capture rather than description bias drives the remaining concentration. The system's output is shaped primarily by the multi-round debate interaction, not by initial framing.

\subsection{Practical Implications}

For policy researchers, the AI Council provides a structured method for surfacing value tensions in policy debates, with built-in quality metrics that indicate which parts of the output to trust. The system does not recommend policies; it maps the landscape of value-based trade-offs, flagging where different perspectives genuinely disagree and where apparent disagreement may reflect poor modeling.

For AI safety researchers, the binary response pattern in small models and the sycophantic capitulation documented in our failed Delphi designs provide empirical data points on the robustness of role-following behavior under adversarial conditions (here, multi-round debate). The finding that architectural heterogeneity dominates prompt engineering as an intervention against unwanted convergence may generalize to other normative domains where maintaining diverse agent behaviors is important.

For industry practitioners building stakeholder simulation tools, model diversity is not optional---it is the primary mechanism for producing useful output.

\subsection{Multi-Agent Architecture vs.\ Single-Model Simulation}

A natural question is whether a single frontier model (e.g., Claude, GPT-4) could achieve comparable results by simulating all seven stakeholders in one inference call. While frontier models can produce plausible stakeholder disagreement, the multi-agent architecture offers three properties that single-model simulation cannot replicate through repeated sampling alone: \emph{genuine architectural diversity} that introduces model-level variation rather than temperature-driven noise around the same central tendency, \emph{statistical characterization} through repeated independent runs whose variation reflects the system's reliability rather than sampling randomness, and \emph{per-evaluator coherence assessment} that flags unreliable voices. A single frontier model run 35 times with temperature $> 0$ would produce distributional data about that model's output distribution---variation around its shared inductive biases---not about the diversity of perspectives that different training regimes encode.

We acknowledge that we do not empirically test this comparison, and the specific contribution of multi-agent heterogeneity versus frontier model capability remains an open question. However, the finding from Fang et al.\ \cite{fang2025ahmad} that heterogeneity does not reduce convergence in accuracy tasks---while it substantially reduces convergence in our normative tasks---provides indirect evidence that the effect is not reducible to ``more models equals more noise'' but reflects genuine differences in how models encode normative positions.

\section{Limitations and Future Work}

We acknowledge several important limitations of this work.

\textbf{Two scenarios.} Our results are based on two policy scenarios that differ in structural properties but do not span the full space of possible policy questions. Generalization to other domains (e.g., environmental policy, economic regulation, health care) requires additional testing.

\textbf{One model pool.} All results use the same seven local models. Coherence scores are a joint measure of model capability and perspective--model fit. Disentangling these would require testing each model across multiple value perspectives---a combinatorial design not feasible with the current seven-model pool but a natural direction for future work. The differentiated profiling results (models scoring high on some perspectives and low on others) suggest that perspective--model fit is a meaningful component, but we cannot rule out that overall model quality contributes to observed coherence differences. This concern is consistent with Li et al.\ \cite{li2025talk}, who find that weak models in heterogeneous multi-agent debate can degrade outcomes---a pattern we observe in the housing scenario where lower-capability models produce the lower-coherence reasoning that drives the fidelity--diversity tradeoff. Different model families, model sizes, or training methodologies may produce different patterns of agreement and disagreement.

\textbf{Fixed champion--option pairing.} Which personality argues for which option is constant across all runs. Varying this pairing could affect evaluator outcomes, as the quality of argumentation may differ across champions.

\textbf{Dominant option in child welfare.} The Intensive option won 25/25 State A runs and 32/35 State B runs. While the system correctly detects this dominance, it cannot create genuine multi-option competition where the underlying models strongly prefer one option.

\textbf{${\sim}$50\% trustworthy tension rate.} Current 8B models maintain faithful value reasoning roughly half the time. This limits the reliability of any individual deliberation, though aggregate patterns across multiple runs are more robust.

\textbf{Frontier model dependency.} The coherence validation layer that improves output quality requires an API call to a frontier model (Claude Sonnet 4), partially undermining the local-first design philosophy of the system. Investigating whether smaller models can serve as reliable coherence validators is a natural next step.

\textbf{No frontier single-model baseline.} We do not compare against asking a single frontier model to simulate all seven stakeholders in one inference call. Such a comparison would help isolate the specific contribution of multi-agent architecture versus model capability. We note that the divergent effects of heterogeneity across task domains; our finding that it reduces convergence in normative deliberation, versus the finding of Fang et al.\ \cite{fang2025ahmad} that it does not reduce convergence in accuracy tasks---provides indirect evidence that the multi-agent normative context produces distinct dynamics, but a direct frontier comparison remains important future work.

\textbf{Five effective perspectives.} Due to tied trait parameters in the Innovator role (Section~3.3), only five of six value dimensions have dedicated evaluators. Creativity is partially but not fully represented, and three of the six theoretical tension pairs involving Creativity are tested only indirectly.

\textbf{Three Security Focus evaluators.} Three of seven evaluators share the Security Focus perspective, structurally weighting vote-based metrics toward this perspective's preferences. While within-state comparisons remain valid (the same structure is present in all states), absolute FCC values and winner distributions reflect this imbalance.

\textbf{Coherence scoring and option--perspective vocabulary overlap.} As noted in Section~4.1, some option--perspective pairings may produce reasoning that maps more naturally onto a perspective's characteristic vocabulary, potentially inflating coherence scores for those pairings independent of reasoning quality.

Future work should extend scenario coverage, test with larger and more diverse model pools, investigate the sensitivity of results to champion--option pairings, and explore the minimum model capability threshold at which Delphi-style self-reflection becomes viable. An additional direction is \emph{reverse advocacy}, where agents argue against their assigned values to generate cross-value framings---a capability enabled by the existing perspective architecture that could automate moral reframing, which humans spontaneously fail to perform in persuasive contexts \cite{feinberg2015moral, feinberg2019moral}.

\section{Conclusion}

We have presented the AI Council, a multi-agent deliberation framework that uses architectural heterogeneity and coherence validation to address artificial consensus in LLM-based policy simulation. Across 120 deliberations on two policy scenarios, we find that using different models for different value perspectives consistently and significantly reduces artificial agreement, with large effect sizes that prompt engineering alone cannot achieve. This finding is specific to normative deliberation, where convergence represents a failure mode rather than a success criterion; architectural heterogeneity disrupts shared inductive biases that drive spurious agreement when no objective ground truth constrains the outcome.

Coherence validation provides a valuable fidelity layer, but its effect on diversity is scenario-dependent---a \emph{fidelity--diversity tradeoff} that we argue is a structural property of any quality-weighted multi-agent system. When weighted and unweighted results diverge, the divergence itself is informative, revealing which value perspectives are modeled reliably and which require stronger underlying models.

More broadly, the two-scenario contrast suggests that multi-agent deliberation systems should be evaluated on structurally diverse problems before deployment. The system runs on consumer hardware, produces interpretable outputs with built-in quality metrics, and preserves disagreement where disagreement exists rather than manufacturing it where it does not.

\section*{Acknowledgments}

The author thanks Maya Arazi for suggesting the application of Delphi methodology to multi-agent validation, which was a key architectural insight. The child welfare scenario is derived from research by Sorek et al.\ \cite{sorek2024child} at the Myers-JDC-Brookdale Institute. All local model inference was performed on consumer-grade hardware using Ollama. Coherence validation used the Claude Sonnet 4 API \cite{anthropic2025claude}.

\appendix

\section{Discrete Values of First-Choice Concentration}

With $N = 7$ evaluators and $K = 3$ options, the maximum number of votes any option can receive ranges from 3 (evenly split) to 7 (unanimous). The normalized first-choice concentration metric maps these to exactly five values:

\begin{table}[t]
\centering
\small
\begin{tabular}{ccc}
\toprule
Max votes & Vote split (example) & FCC \\
\midrule
3 & 3-2-2 & 14.3\% \\
4 & 4-2-1 & 35.7\% \\
5 & 5-1-1 & 57.1\% \\
6 & 6-1-0 & 78.6\% \\
7 & 7-0-0 & 100.0\% \\
\bottomrule
\end{tabular}
\end{table}

This discrete structure means that small changes in voting can produce large jumps in concentration, and that intermediate values (e.g., 50\%) are impossible. All reported means are averages over these discrete values.

\section{Rebalanced Scenario Test}

To test whether the dominance of the Intensive option in the child welfare scenario was driven by asymmetric option descriptions (Intensive describes more resources), we rewrote all three descriptions to be more balanced in appeal. The rebalanced scenario produced first-choice concentration of 48.6\%, compared to a baseline range of 46--49\% in the original scenario. This null result confirms that the remaining concentration in State B is driven by debate dynamics (argument quality, persuasive debate interactions) rather than initial description bias.

\end{document}